\documentclass{article}
\usepackage{spconf,amsmath,graphicx,hyperref}
\usepackage{booktabs}
\usepackage{bbding}
\usepackage{xcolor}
\usepackage{soul}

\ninept

\title{Contextual biasing for ASR in speech LLM with \\ common word cues and bias word position prediction }

\name{Sashi Novitasari, Takashi Fukuda, Gakuto Kurata, George Saon}
\address{IBM Research}

\begin{document}

\maketitle

\begin{abstract}

Speech-aware LLMs (SLLMs) have recently achieved \textit{state-of-the-art} ASR performance; however, they still fail to accurately transcribe bias words that appear rarely or never in the training data. 
Contextual biasing mechanisms are commonly implemented by introducing a predefined bias word list into the model via a text prompt or additional module. For further improvement, predefined bias words can be paired with their phoneme representations as pronunciation cues. Typically, phoneme sequences are generated through a G2P system that covers the target languages and domains of the bias words. Therefore, when a compatible G2P system is unavailable, phoneme-assisted contextual biasing becomes difficult to perform. Moreover, manually adding accurate phoneme sequences requires advanced phonetic knowledge. In this paper, we explore contextual biasing in SLLM based on acoustic cues associated with a set of common words whose pronunciations are partially similar to those of the target bias words. We assume ASR applications in which end users do not require special knowledge of phonetics or utilize G2P tools for inference. For enhanced robustness, we also introduce bias word positional prediction implemented in a multi-output learning fashion. Our method reduces bias word recognition errors by 16.3\% compared to baseline systems, including on out-of-domain data.
\end{abstract}


\vspace{-0.3cm}
\section{Introduction}
\vspace{-0.2cm}

\label{sec:intro}
Automatic speech recognition (ASR) technology has advanced rapidly in recent decades.  
The latest developments have enabled ASR in large language models (LLMs) to achieve \textit{state-of-the-art} performance. 
In particular, speech-aware text LLM (SLLM) frameworks \cite{ma_2024_slamllm,grattafiori_2024_llama3,microsoft_2025_phi4,saon_2025_granitespeech} have gained attention for their remarkable modularity and performance. 
These frameworks augment a text LLM with a speech encoder, enabling audio-based tasks while preserving the core text-based capabilities. 
In spite of the great success on end-to-end ASR modeling, ASR in SLLMs still faces challenges in accurately transcribing bias words, which are words that are underrepresented in the training data (i.e., rare or unseen words). 
Due to limited training examples, these words may be either deleted or substituted with other phonetically similar words that may appear more frequently in the training data. 
Previously, non-LLM ASR systems employed contextual biasing mechanisms to accurately transcribe bias words by deep-fusing a predefined list of bias words into the ASR model \cite{jain2020,tang_2024,yolwas2025adaptive}. 
In SLLM research, prompt-based contextual ASR approaches \cite{sun_2023,gong24b_interspeech,yang_2024,he_2025cmt} have recently gained attention by incorporating a bias list into the text prompt alongside task instructions.
These approaches leverages the SLLM text module to directly process bias list without requiring additional bias encoder modules. 
To handle long bias lists, previous studies have also proposed list pruning methods \cite{gong_2025,hou25_interspeech}.

Although contextual biasing enhances the model's ASR capability, errors still occur on words with unusual spellings or pronunciations. 
To address this problem, previous studies \cite{Pandey_2023,qiu_2023,he_2025} decoded the ASR output by supplementing the predefined bias words with their pronunciation hints, represented as phonemes generated via a grapheme-to-phoneme (G2P) system before inference. 
Despite the potential performance gains, phoneme-assisted systems face challenges in real-world situations.
The primary issue arises when the target bias word is beyond the scope of the G2P systems. 
This is because typical G2P systems are often designed for specific domains, languages, and phoneme formats. 
For systems with a dictionary-based G2P, users must manually add a new entry of bias word with its pronunciation.
While the user-provided pronunciations are expected to be accurate, this task requires linguistic expertise that is uncommon among end users. 
As a result, the provided pronunciations can be inaccurate or not supplied by users due to the complexity of the task, thus limiting the effectiveness of phoneme-assisted contextual biasing.
Even in systems with more sophisticated G2Ps, the estimated phonemes for rarely-pronounced bias words may still be inaccurate due to domain mismatch.
Addressing such mismatches often requires model retraining, which leads to higher development costs than those of standard contextual ASR model that do not rely on explicit phonetic assistance. 
Furthermore, integrating any type of G2P system, including a neural-based one, also increases system complexity during inference, which is undesirable in low-resource scenarios.

In this work, instead of using direct phonetic representation, we explore common or non-bias words with phonetic similarity to bias words as pronunciation cues or hints for the bias words (e.g., ``\textit{Shelley}" - ``\textit{sheriff, legal"}).
Common words appear frequently in large-scale training data; therefore, SLLM possesses sufficient knowledge of their pronunciation, making them useful as pronunciation reference for bias words. 
The proposed cues can be generated manually or via simple programming without any advanced knowledge of phonetics; users may create it based on pronunciations derived from a set of more commonly known words.
If a G2P system is applicable during inference, the proposed cues can also be generated using any G2P system flexibly without needing to re-adapt the SLLM, even when the G2P system is not the same as the model utilized during training.
For robust hint generation, our method assigns hints on the basis of partial phonetic match between common word and target bias word, since common word with full phonetic matches are often difficult to find, especially for phonetically complex words.
In this paper, we explore several criteria for selecting hint words based on their phonetic and spelling similarities to the target bias word.
To the best of our knowledge, there is little prior research  employing word-level cues for contextual biasing tasks. 
We assume user experience with SLLMs can be enhanced through the proposed acoustic cues.

To maintain the broad applicability of the SLLM for ASR tasks, we ensure model robustness across three types of tasks simultaneously: (1) basic non-contextual ASR task, (2) contextual ASR task without the pronunciation hints, and (3) contextual ASR task with the proposed hint, and without increasing computational costs during inference. 
To this end, we also propose a multi-output training mechanism by augmenting the SLLM with an additional removable module for predicting a bias word positional tag based on the latent features produced by the speech encoder projector and LLM. 
This mechanism is applied while training the model by using the prompts for the target tasks within a single training pipeline.
The proposed method aims to encourage the model to better distinguish the audio or text transcription parts that belong to the bias or the non-bias word, thus, improving the overall word accuracy.
The additional module for the bias word positional tagging is removed during inference; therefore, the proposed SLLM structure remains unchanged from the standard structure consisting of a speech encoder, projector, and text LLM.
In summary, our contributions are as follows:
\begin{enumerate}
    \item We propose word-level cue representations based on common words as pronunciation hints for bias words with high applicability.
    \item We demonstrate  word-level cue selections based on phonetic (pronunciation) and structural (spelling) similarities between the common word and the bias word.
    \item We apply an SLLM training framework with a bias word position prediction mechanism to improve the model's generalization while leveraging hint-assisted contextual ASR. 
\end{enumerate}

\vspace{-0.3cm}
\section{Methodology}
\vspace{-0.2cm}

\subsection{Contextual biasing via textual prompt}
\vspace{-0.2cm}

Our model employs a textual prompt-based approach to perform contextual biasing on ASR tasks.
It takes input consisting of a speech audio $S=[s_1,s_2,...,s_I]$ with $I$ frames, a textual task instruction $X=[x_1,x_2,...,x_J]$ with $J$ words (e.g., "Transcribe this speech"), and  a bias word list $B=[b_1,b_2,...,b_K]$  with $K$ bias words. 
$X$ and $B$ are concatenated into a single text before being fed to the SLLM.
The model output is a speech transcription $T=[t_1,t_2,...,t_L]$ consisting of $L$ words, where $0 \leq |B \cap T| \leq L$ words. The overall operation can be expressed as $T = \text{SLLM}(S,X,B)$. 

\vspace{-0.2cm}
\subsection{Proposed contextual biasing with common word cues}
\vspace{-0.2cm}
The phonetic cues for bias words are provided to the SLLM via text prompts, where they are paired with the corresponding bias words in $B$.
As the primary method, we use phoneme as pronunciation cues for bias words by assuming an ideal situation where high-quality phoneme input is available for training and inference.
For our proposed method, we explore several criteria below to select the hint words based on phonetic and structural similarities to the bias word.

\vspace{-0.2cm}
\subsubsection{Syllable-based partial phonetic match} 
\vspace{-0.2cm}

The pronunciation hint for each bias word is a sequence of common words whose first syllables, when spoken in sequence, resemble the target bias word.
In the text prompt, a bias word $b_k$ that has $M$ syllables ($b_k=[sb_{k_1},...,sb_{k_M}]$) is paired with a set of other $M$ words ($H_k=[h_{k_1},...h_{k_M}]$) as its hint.
The $m$-th hint word is composed of $R$ syllables ($h_{k_m}=[s_{km_1},...s_{km_R}]$), whose first syllable $s_{km_1}$ has the same or similar phoneme sequence as $sb_{k_m}$, the $m$-th syllable in $b_k$.
This approach decomposes the phonetic word matching problem into partial matching, enabling intuitively more tractable hint representation than a raw phoneme sequence. 
We specifically use the first syllable as the matching criterion to simplify the SLLM's task for identifying the bias word's pronunciation from the hints.

In our experiments, the proposed hint is generated through two steps for each bias word syllable $sb_{k_m}$. Given $sb_{k_m}$, we first retrieve all words from a common word list constructed using a public or in-house word list, or both, where the retrieved word's first syllable has a similar phoneme sequence to that of $sb_{k_m}$.
Second, we select a word from the retrieved word candidates as the final hint word $h_{k_m}$.
Here, we explore two approaches to select $h_{k_m}$: selection by the smallest character-level edit distance (CED) to the target bias word, and random word selection. The CED-based selection produces hint word with spelling or structural similarity to the corresponding bias word. 
Meanwhile, random selection simulates a variety of user-provided manual inputs as hints. 
In our experiments, CED-based word selection was applied during training and inference, while random selection was performed only during inference to evaluate model robustness on various candidates of hint words.

\vspace{-0.1cm}
\subsubsection{Phonetic vowel match} 
\vspace{-0.2cm}
In this approach, a bias word's cues consist of a sequence of words that constructs a similar phonetic vowel pattern to that of the target word.
For example, a bias word ``\textit{Shelley}'' can be paired with a hint word ``\textit{healthy}'' for sharing the same vowel sequence ``\textit{EH-IY}".
We choose vowel-based matching because vowels play an important role in determining how a word sounds. 
Formally, bias word $b_k$ is paired with a word sequence $H_k=[h_{k_1},...h_{k_U}]$ consisting of $U$ words.
The sequence length $U$ can be one word if a common word with an exact vowel match to the target word is available. 
When such words are unavailable, $H_k$ is extended with other common words to match the target vowel sequence.
In our experiments, the final $H_k$ sequence is also selected based on CED or random selection when multiple hint candidates are available.
Similar to section 2.2.1, we assume the CED criterion is applied mainly at training time, while the random selection approach is explored at inference time to simulate various user inputs.

\vspace{-0.2cm}
\subsubsection{Closest character-level and phoneme-level edit distance}
\vspace{-0.2cm}

In this method, the predefined bias word $b_k$ is paired with a hint word $h_k$ from the viewpoint of the smallest CED out of other common words, without relying on syllable-based matching.
Here, we examine the effectiveness of word structural similarity in selecting hint words.
Since multiple common words may simultaneously have the smallest CED to the bias word, we also investigated different selection methods to choose $h_k$.
During model training, the final $h_k$ is selected  based on phoneme-level edit distance (PED) to the target word when multiple candidates have the smallest CED.
For inference, we also explored a random selection approach among the multiple CED-selected candidates in our experiments.

\vspace{-0.2cm}
\subsection{Proposed bias word position prediction}
\begin{figure}[t]
  \centering
  \includegraphics[width=0.84\linewidth]{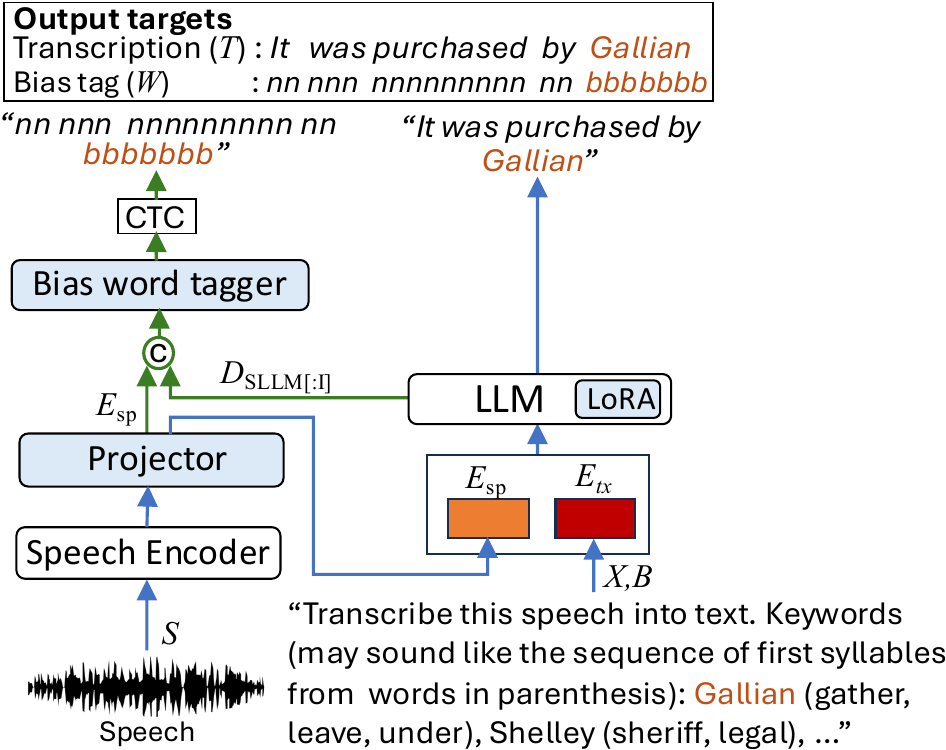}
    \vspace{-0.4cm}
    \caption{The proposed SLLM training with a bias word position tag prediction mechanism. The bias list applies the proposed cues using syllable-based partial phonetic matching. (``\textit{b}'': character-level bias word tag, ``\textit{n}'': character-level non-bias word tag) }
  \label{fig:overw.frmwork}
  \vspace{-0.35cm}
\end{figure}

\vspace{-0.2cm}
To enhance the SLLM's robustness on contextual ASR tasks without degradation on non-contextual ASR tasks, we train our model through a multi-task and multi-output training framework. 
Multi-task training is performed by training the SLLM on prompts of multiple target tasks together. 
On the other hand, the proposed multi-output training mechanism (Fig. \ref{fig:overw.frmwork}) trains the SLLM to predict the positions of bias words within an utterance, in addition to speech transcription. 
The bias word positions are represented as a sequence of character-level tags  $W=[w_1,w_2,...,w_V]$ consisting of $V$ tokens, where each is associated with a character in the speech transcription. Our tag set consists of the ``\textit{bias}", ``\textit{non-bias}", and ``\textit{whitespace}" tags. The ``\textit{bias}" tag  is assigned to the character sequence of bias word, and likewise for other categories.

The bias word position is tagged on the basis of two features: (1) the latent features produced by speech encoder projector $E_{sp}=[e_1,e_2,...,e_I]$ with a length of $I$ frames, and (2) the LLM causal output $D_{LLM}$ for the past states that align with $E_{sp}$. 
These features are concatenated element-wise before being fed to the bias word tagger. The overall operation can be expressed as:
\begingroup
\setlength\abovedisplayskip{2pt}
\begin{equation}
        D_{LLM} = \text{LLM}(E_{sp},E_{tx}),
                \vspace{-0.2cm}
\end{equation}
\endgroup
\begingroup
\begin{equation}
        W = \text{Bias word tagger}(E_{sp},D_{LLM[1:I]}),
        \vspace{-0.1cm}
\end{equation}
\endgroup
where $E_{tx}$ is the text prompt embedding and $D_{LLM[1:I]}$ aligns with $E_{sp}$.
The bias word tagger is optimized by using the Connectionist Temporal Classification (CTC) \cite{graves_2006} loss; thus, the bias tag sequence $W$ also corresponds to the speech frame sequence. 
The proposed model's loss formulation ($L_{SLLM}$) is as follows: 
\begingroup
\setlength\abovedisplayskip{4pt}
\begin{equation}
        L_{SLLM} = L_{ASR}(T,\hat{T}) + \alpha L_{CTC}(W,\hat{W}), 
        \vspace{-0.3cm}
\end{equation}
\endgroup
where $L_{ASR}$ is the loss for the speech recognition task, $\hat{T}$ is the speech transcription output, $\hat{W}$ is the bias word positional tag output, and $\alpha$ is the loss coefficient for bias word tagger.

\vspace{-0.2cm}
\section{Experiment setting}
\vspace{-0.3cm}

\subsection{Model}
\vspace{-0.2cm}

We used the Granite-Speech \cite{saon_2025_granitespeech}\footnote{https://huggingface.co/ibm-granite/granite-speech-3.3-8b} architecture as our SLLM backbone, which was originally designed for ASR and speech translation tasks. 
In our experiments, we focus on English ASR tasks. 
It consists of a speech encoder, a projector, and a text LLM. 
The speech encoder has a Conformer-CTC structure with 10 Conformer blocks, which was pretrained for a character-level ASR task. The speech encoder is connected to a Q-former \cite{li_2023_qformer} projector that downsamples and projects the speech encoder output into the LLM embedding space. We used the \textit{granite-3.3-8b-instruct} model \footnote{https://huggingface.co/ibm-granite/granite-3.3-8b-instruct} as the text LLM. The contextual biasing capability was added to our SLLM through a post-training mechanism, in which we fine-tuned the model on the ASR dataset using contextual ASR prompts. 
Our model was tuned through three epochs with a learning rate of 5e-6. Only the Q-former and LoRA \cite{hu_2021_lora} parameters associated with the LLM were updated. 
For the proposed bias word tagger, our best setting employed nine FNN layers for models with syllable or vowel-based hints, and eleven layers for models with CED+PED-based hint.

\begin{table}[t]
\centering
\caption{The frequency of bias and non-bias words in test data.}
\resizebox{0.36\textwidth}{!}{
\begin{tabular}{l|c|c}
\hline
\multicolumn{1}{c|}{Test data}           & \#Bias word   & \#Non-bias word \\
\hline
Librispeech test-other  & 2,050 & 50,832 \\
Common voice & 15,669 & 129,456 \\
SPGI &  12,405 & 956,711 \\
Gigaspeech  &  15,520 & 388,134  \\
\hline
\end{tabular}}
\label{tbl:bias_word_count}
\vspace{-0.6cm}
\end{table}

\bgroup
\def\arraystretch{1}%
\setlength{\tabcolsep}{1pt}
\begin{table}[t]
\caption{Contextual ASR performance (\%) of the proposed SLLM on Librispeech \textit{test-other}. The bias list size was 200 words. (``\textit{Non-ctx}'': non-contextual ASR, ``\textit{Ctx}'': contextual ASR, ``\textit{Phon}": phoneme, ``\textit{Syl}'': syllable-based partial phonetic match, ``\textit{Phon.vow}'': phonetic vowel match, ``\textit{rand}": random selection.)}

\resizebox{0.49\textwidth}{!}{
\begin{tabular}{l|l|l|ccc}
\hline
\multicolumn{2}{l|}{Bias hint selection criteria} & Hint & B-WER & U-WER & WER \\ \cline{1-2}
Train prompt                    & Test prompt       & type &       &       &     \\
 \hline \hline
 1. Baseline Non-ctx               & Non-ctx           & -  &  20.5&	2.3 & 3.0  \\
 2. Baseline Ctx,no phonetic hint  & Ctx,no phonetic hint  & -  & 5.8 & 2.2 & 2.3 \\
3. Topline Ctx, Phon               & Ctx, Phon              & Phon & 3.4 & 2.2 & 2.2 \\
 \hline
 \multicolumn{6}{l}{\textbf{Ctx with the proposed word-level cues}} \\
 4. Syl+CED                        & Syl (rand)        & Word & 5.1 & 2.2 & 2.3 \\
                                & Syl+CED           & Word & 5.1 & 2.2 & 2.3 \\
\hline
 5. Phon.vow+CED                   & Phon.vow (rand)  & Word & 5.4 & 2.1  & 2.3 \\
                                & Phon.vow+CED     & Word & 5.3 & 2.2  & 2.3 \\ 
                                
\hline
6. CED+PED                         & CED (rand)        & Word & 4.4 & 2.1 & 2.2 \\
                                & CED+PED           & Word & 4.4 & 2.1 & 2.2 \\    

 \hline
\end{tabular}}
\label{tbl:lib_ne_200}
\vspace{-0.6cm}
\end{table}
\egroup

\bgroup
\def\arraystretch{1}%
\setlength{\tabcolsep}{2pt}
\begin{table}[t!]
\centering
\caption{Contextual ASR performance (\%) of the proposed SLLM on Librispeech \textit{test-other}. The bias list size was 10 words.}
\resizebox{0.36\textwidth}{!}{
\begin{tabular}{l|c|ccc}
\hline 
\multicolumn{1}{l|}{Bias hint  selection criteria} & Hint  & B-WER & U-WER & WER \\ 
\multicolumn{1}{l|}{(Train and test prompts)} & type &  &  & \\ \hline \hline
1. Ctx, no phonetic hint           & -  & 4.2 & 2.1	 & 2.2  \\
2. Ctx, Phon                       & Phon & 2.3 & 2.1	 &	2.1 \\
 \hline 
 \multicolumn{5}{l}{\textbf{Ctx with the proposed word-level cues}} \\
3. Syl+CED          & Word & 3.8  &	2.1 &	2.2 \\
4. Phon.vow+CED      & Word & 3.2 &	2.1 &	2.2	 \\
5. CED+PED           & Word & 3.2 &	2.1 &	2.2 \\
 \hline
\end{tabular}}
\label{tbl:lib_ne_oracle}
\vspace{-0.5cm}
\end{table}
\egroup

\bgroup
\def\arraystretch{1}%
\setlength{\tabcolsep}{3pt}
\begin{table*}[t]
\centering
\caption{
ASR performance (\%) on different ASR tasks. The proposed SLLMs were trained through the multi-output mechanism. The bias list size was 200 words. Results with the same ID came from the same model using different inference prompts.}
\resizebox{0.64\textwidth}{!}{
\begin{tabular}{c|l|cc|cc|cc||cc}
 \hline
ID &\multicolumn{1}{c|}{Model} & \multicolumn{2}{c|}{Common voice} & \multicolumn{2}{c|}{SPGI} & \multicolumn{2}{c||}{Gigaspeech} & \multicolumn{2}{c}{Avg.} \\ \cline{3-10} 
 & \multicolumn{1}{c|}{} & B-WER & U-WER  & B-WER & U-WER & B-WER & U-WER & B-WER & U-WER \\ \hline \hline
& &\multicolumn{8}{l}{\textbf{Inference: Non-contextual ASR (no bias list)}} \\
1. &Non-ctx  & 22.6 &5.5&		15.6&	3.0&	27.2&	9.8&	21.8&	6.1 \\
2. &Ctx, no phonetic  hint  &  23.0 &5.7 &15.9 &3.2 &26.7 &9.5 &21.9 &6.1 \\ \hline
3. & Syl+CED & 23.0 &5.8 &16.1 &3.3 &26.8 &9.6 &22.0 &6.2 \\ 
4. & Phon.vow+CED   &23.2 &5.8 &16.0 &3.3 &27.1 &9.5 &22.1 &6.2\\ 
5. & CED+PED & 23.0 &	5.8&		16.1&	3.3&	26.8&	9.6&	22.0 &6.2 \\
\hline \hline
& & \multicolumn{8}{l}{\textbf{Inference: Standard contextual ASR (no phonetic hint for bias words)}} \\
2. &Ctx, no phonetic hint &9.2 &5.5 &5.2 &3.2 &17.3 &9.6 &10.6 &6.1 \\ \hline
3. & Syl+CED & 8.9 &5.5 &5.2 &3.3 &16.9 &9.6 &10.3 &6.1 \\
4. & Phon.vowel+CED  &9.3 &5.5 &5.2 &3.3 &16.8 &9.5 &10.4 &6.1 \\
5. & CED+PED & 9.0&	5.6&		4.9&	3.2&	16.7&	9.5&	10.2 &6.1 \\
\hline \hline
& & \multicolumn{8}{l}{\textbf{Inference: Contextual ASR with the proposed word cues}} \\
3. & Syl+CED   & 7.6 &5.6 &4.3 &3.3 &16.0 &9.5 &9.3 &6.1 \\
4. & Phon.vowel+CED   & 8.1 &5.5 &4.4 &3.3 &15.9 &9.4 &9.4 &6.1 \\
5. & CED+PED & 7.0&	5.5&		3.9&	3.3&	15.7&	9.5&	8.8 &6.1\\
\hline
\end{tabular}}
\label{tbl_mtask_main}
\vspace{-0.7cm}
\end{table*}
\egroup

\bgroup
\def\arraystretch{1}%
\setlength{\tabcolsep}{3pt}
\begin{table}[]
\centering
\caption{Comparison of B-WER (\%) on Common voice data between the models trained with the
single-output (transcription only) and the proposed multi-output mechanism.}
\resizebox{0.36\textwidth}{!}{
\begin{tabular}{l|c|c|c}
\hline
Model (Syl+CED) & \multicolumn{1}{c|}{Non-ctx} & \multicolumn{1}{c|}{Ctx (no hint)} & \multicolumn{1}{c}{Ctx+hint} \\ \hline
\phantom{qq}Single-output       &   23.2 &	9.3 &	8.3 \\ 
\phantom{qq}Multi-output      &  23.0  &	8.9  &	7.6 \\ \hline
\end{tabular}}
\label{tbl:mout_sout}
\vspace{-0.7cm}
\end{table}
\egroup

\vspace{-0.3cm}
\subsection{Dataset}
\vspace{-0.2cm}

The training datasets described in this section were used to add contextual ASR capability to the backbone SLLM.
In our initial experiment, we trained our models using the Librispeech \cite{panayotov_2015_ls} corpus as the basic setting to evaluate the proposed bias word cues.
In the second experiment, we assessed our complete proposed pipeline on a larger data scale. The training corpora consisted of Librispeech, CommonVoice 17.0 \cite{ardila_2020_cv}, Voicemail \cite{padmanabhan_2002_voicemail}, AMI \cite{mccowan_2005_ami}, and Voxpopuli \cite{wang_2021_voxpopuli}.
Evaluations were conducted on CommonVoice 17.0 as an in-domain setting.
To assess robustness, we also tested our models on out-of-domain datasets, SPGI \cite{oneill_2021_spgi} and Gigaspeech \cite{chen_2021_gs}, which cover a  wide variety of acoustic signals and linguistic topics.
The bias list in all experiments was constructed by automatically extracting named entities from the speech transcriptions using a named entity tagger. 
Table \ref{tbl:bias_word_count} shows the word statistics of our test data.

For the proposed methods described in Section 2.2, we first converted the bias words and common words into their phoneme sequences to perform  phonetic matching. 
The common word list for word-level cues was constructed using the MIT 10K word list \footnote{https://www.mit.edu/$\sim$ecprice/wordlist.10000}, excluding the target bias words. 
Words were converted into phoneme sequences using a manually-labeled word dictionary that defines phoneme sequences for each word and the public SoundChoice G2P model \cite{Ravanelli_2021_speechbrain,ploujnikov_2022_soundchoice}.
The bias list size for training was randomized between one and 200 words for each utterance. During inference, bias lists of ten or 200 words were used, containing all bias words of the corresponding utterance and random distractors.

\vspace{-0.2cm}
\section{Results and discussion}
\vspace{-0.3cm}
\subsection{SLLM with proposed word cues for bias words}
\vspace{-0.2cm}

First, we independently investigate the impact of the proposed word-level acoustic cues for contextual biasing without the proposed multi-output training, based on the models trained only on the Librispeech dataset. 
Table \ref{tbl:lib_ne_200} shows the models' performances on bias list of 200 words, while Table \ref{tbl:lib_ne_oracle} shows results on the shorter bias list that consisted of ten words.
We employed three word error rates (WERs) commonly used in conventional works to evaluate our models:
B-WER that considers only the bias words and removes the non-bias words from the transcription, 
U-WER that considers only the non-bias words, and the full WER based on the entire transcription. 
Our baseline was the contextual ASR-trained SLLM that used the predefined bias list but without the phonetic hint (``Ctx, no phonetic hint''), 
while the topline model used phonemes as the bias word's hint (``Ctx-Phon''), which is an oracle case in our experiments.
As expected, models equipped with contextual ASR capability had significantly lower B-WER than the model without contextual biasing capability (``Non-ctx"), while U-WERs were maintained.
Full WER was less affected by B-WER because the frequency of bias words was significantly lower than that of non-bias words.

As shown in Table \ref{tbl:lib_ne_200}, 
our proposed models demonstrated relatively significant B-WER reduction compared to the baseline (2) by up to 24.1\% (B-WER=5.8\% to 4.4\%), narrowing the gap to the oracle model (3) in our assumption. 
These results imply that the proposed hint representations could enable the SLLM to transcribe bias words more accurately.
The performance gap relative to the oracle model (3) reflects a trade‑off for reduced  user effort or complexity.
Overall, the lowest B-WER was achieved using the proposed ``CED+PED" model trained using the cues selected only by CED and PED. 
It also performed robustly when using the CED-only selected cues without considering PED at inference time, but with a random selection.
This may be because the cues consisted of a single word, which was more concise than the other type of cues (sequence of words), making them easier for the SLLM to process. 
Interestingly, the proposed ``Phon.vow+CED" model, which used cues from phonetic vowel matching, resulted in the lowest B-WER with a short bias list, but it did not achieve the same when more distractors were included into the list. 
Since the cues were only selected  based on phonetic vowel similarity,
distractors may also have been paired with hint words similar to those of relevant bias words, potentially confusing the model as the number of distractor increased.
Meanwhile, the proposed ``Syl+CED'' model, which was trained using cues based on syllable similarity, resulted in the second-highest performance.
The proposed syllabically-selected cues provided richer and more relevant cues and represent intuitively the simplest generation method if provided manually.
All proposed models also performed robustly when random selection was applied to select the final hint sequence instead of edit distance-based selection, leading to more user-friendly systems.

\vspace{-0.3cm}
\subsection{SLLM with proposed bias word position prediction}
\vspace{-0.2cm}

In this experiment, we investigated the proposed models trained on the full-scale dataset through the multi-output approach described in Section 2.3. Experimental results on different ASR prompts during inference are shown in Table \ref{tbl_mtask_main}.
To simulate novice user input without expertise of phonetics, the final hint words at inference time were randomly selected when multiple candidates were available; the bias hints for ``Phon.vow+CED" and ``Syl+CED" models were selected solely on the basis of phonetic similarity.

Our results show that our proposed approaches improve the SLLM's generalization across multiple ASR tasks.
All proposed models outperformed the baseline model on the standard contextual ASR task (no hint), indicating that the proposed cues also enhanced the model's  learning process for identifying the bias word correctly, even for the situations where no phonetic cues are provided.
When word-level phonetic cues were applied during inference, the proposed models yielded average  performance gains of 11.3\%-16.3\% relative to the baseline (B-WER=10.6\% to 9.4\%-8.8\%). 
Additionally, as shown in Table \ref{tbl:mout_sout}, we investigated the performance of the proposed model with and without the proposed auxiliary task. 
The proposed multi-output approach improved contextual ASR B-WER (with and without hints) by up to 8.4\%, demonstrating that the proposed method enhanced the SLLM's ability to correlate bias words spoken in the speech with those hinted through the text prompt.

\vspace{-0.3cm}
\section{Conclusion}
\vspace{-0.3cm}
We proposed a contextual ASR method for SLLM using common words as phonetic cues for bias words and multi-output training with bias word positional prediction.
Our results demonstrated that the proposed word-level cues enhanced the contextual ASR performance in SLLM, while the proposed multi-output training method also improved the model's generalization. 
The proposed methods yielded consistent performance gains across multiple ASR tasks, supporting a robust and versatile model applicable to a wide range of ASR scenarios.

\bibliographystyle{IEEEbib}
\bibliography{strings,refs}

\end{document}